\documentclass[aps,twocolumn,groupedaddress,superscriptaddress,showpacs,prb,floatfix]{revtex4}
\usepackage{graphicx}% Include figure files
\usepackage{dcolumn}% Align table columns on decimal point
\usepackage{bm}% bold math
\usepackage{epsfig}
\usepackage{dsfont,psfrag}\usepackage{color}
\usepackage{ifthen}
 \newcommand{\ket}[1]{{| #1\rangle}}

\begin{document}
\bibliographystyle{amsplain}
%  \title{Quantum state transfer in a homogeneous XX chain induced by boundary magnetic fields}
  \title{Perfect state transfer in XX chains induced by boundary magnetic fields}
\author{Thorben Linneweber}
\author{Joachim Stolze}
\email{joachim.stolze@tu-dortmund.de}
\author{G\"otz S. Uhrig}
\affiliation{Technische Universit\"at Dortmund, Institut f\"ur Physik, 
D-44221 Dortmund, Germany}

\date{\today}
\begin{abstract}
A  recent numerical study of short chains found near-perfect quantum state
transfer between the boundary sites of a spin-1/2 XX chain if a sufficiently 
strong magnetic field acts on these sites. We show that the phenomenon is 
based on a pair of states strongly localized at the boundaries of the system 
and provide a simple quantitative analytical explanation.
\end{abstract}
\pacs{03.67.Hk,03.65.Yz,75.10.Pq,75.40.Gb}
\maketitle

\section{Introduction}
\label{one}
The transfer of quantum states between different elements of a quantum
computer is an important and 
%demanding 
challenging task in quantum information
processing. Chains of interacting spin-1/2 particles were proposed 
as a means to achieve this task. Many different types of quantum spin
chains have been discussed, as evident, for example, from the reviews 
\onlinecite{Bos07} and \onlinecite{Kay10}. 
In the most
frequently studied scenario the first site of a finite spin chain is
prepared in a definite state which is then supposed to be transferred
to the last site by the natural Hamiltonian dynamics. 

Since the
elementary excitations of spin chains are often dispersive, the
initially sharply localized quantum state broadens as it propagates
and thus quantum information degrades \cite{OL04}  so that the state received
 is not identical to the state transmitted. In view 
 of limits to fault tolerance in quantum information processing it is,
however, desirable to achieve {\em perfect state transfer} (PST). Engineered
chains were shown to be attractive means in reaching this goal 
\cite{CDEL04,ACDE04,YB04,SLSS04,my44,Kay10,WSR11}.
In these chains, all spin-spin couplings and / or local
fields are set at certain values which guarantee PST. 

However, from a practical point of view, it would be more
desirable to achieve PST by simpler means, for example by manipulating
only ${\cal O}(1)$ system parameters instead of the  ${\cal O}(N)$
coupling constants which must be adjusted for an $N$-site engineered
PST chain. This line of thought has been pursued in a number of
studies in which the ends of a homogeneous spin chain, i.e., one with
constant couplings, were connected to a single transmitting and
receiving qubit, respectively, by adjustable coupling constants 
\cite{WLKx05,BACx10,BACx11,ZO11,OWSx11,YJGx11} which influence the 
transfer of states via different mechanisms, depending on other details 
of the systems employed. 

In a similar vein, Casaccino et
al. \cite{CLMS09} suggested to use sufficiently strong magnetic fields acting
exclusively on the boundary spins of an otherwise homogeneous spin
chain. In their numerical results, they indeed observed near-perfect
state transfer, however, with the transfer time growing strongly with
both the chain length and the magnetic field strength.

In this paper, we explain the effect observed in Ref. \onlinecite{CLMS09} by
analytical considerations. We show that boundary magnetic fields of
sufficient strength $h$ produce exceptional energy eigenstates of the
chain. These states are exponentially localized at the boundaries with
localization length $\sim (\ln h)^{-1}$. The PST is completely
determined by these states and the dynamically relevant energy
differences become exponentially small as the system size
grows. Correspondingly the time scales become exponentially large,
making this kind of PST unfortunately impractical for long
chains. Furthermore, since the PST depends on an exponentially {\em
  weak link} between transmitter and receiver (the central region of
the chain, where the amplitudes of the relevant states are
exponentially small) one might speculate that any slight imperfection is 
bound to strongly
affect the PST performance. The robustness of state transfer against disorder 
in the couplings was recently studied for several kinds of both engineered and 
boundary-controlled spin chains \cite{my60,my62}.

In Sec. \ref{two} of this paper we present analytical solutions
for the energy eigenstates and eigenvalues of the homogeneous semi-infinite 
chain with
a magnetic field at the first site. The results thus obtained suggest
a simple perturbative calculation and an approximate
expression for the exact result for the finite chain with boundary
fields at both ends, which we discuss in Sec. \ref{three} and which
explains the numerical observations in Ref.\ \onlinecite{CLMS09}.

\section{The semi-infinite chain: exact results}
\label{two}

The spin-1/2 XX chain with boundary magnetic fields is defined by the 
Hamiltonian
\begin{equation}
  \label{eq:1}
  H = \sum_{i=1}^{N-1} 2J 
\left(
S_i^x S_{i+1}^x + S_i^y S_{i+1}^y 
\right)
+ h 
\left(
S_1^z + S_{N}^z
\right),
\end{equation}
where $S_i^x$,  $S_i^y$, and $S_i^z$ 
are the usual spin-1/2 operators with eigenvalues
$\pm 1/2$. The most popular scenario in quantum information transport
studies uses an initial product state, $\ket{\psi(t=0)} = \ket{\varphi}
\bigotimes_{i=2}^N \ket{\downarrow}_i$, with $\ket{\varphi} = \alpha
\ket{\uparrow} + \beta \ket{\downarrow}$, i.e., a superposition of the
state with no  spin up and a state with one single  spin up located at site 1.
 Since the
Hamiltonian conserves the number of spins `up' and since the
zero spin up component does not evolve in time it suffices to follow
the evolution of the single spin up component. The corresponding
$N$-dimensional subspace is spanned by the states $\ket{i}=S_i^{+}
\ket{\downarrow \downarrow...\downarrow}$ and the 
Hamiltonian matrix
$H_{ij}$ is given by
\begin{equation}
  \label{eq:2}
  H_{i,i+1} = H_{i+1,i}= J ; H_{11}=H_{NN}=h .
\end{equation}
For
$h=0$ the eigenvectors $\vec u^{\nu}$ of the matrix (\ref{eq:2}) are
given by their components 
\begin{equation}
  \label{eq:3}
  u_l^{\nu} = 
\left(
\frac 2{N+1}
\right)^{\frac 12}
\sin \frac {\nu \pi l}{N+1}
\end{equation}
with $\nu=1,...,N$, and the energy eigenvalues are 
\begin{equation}
  \label{eq:4}
  E_{\nu} = 2 J \cos \frac {\nu \pi }{N+1}.
\end{equation}
By the Hellmann-Feynman theorem \cite{Gue32,Pau33,Hel37,Fey39} 
all eigenvalues grow if a
magnetic field $h>0$ is switched on. To see more precisely what
happens for $h \neq 0$ we  now consider the case $N \to \infty$,
i.e., we  effectively study a semi-infinite chain with boundary
magnetic field $h$. Since we are looking for states localized near the
boundary $l=1$ we use the Ansatz
\begin{equation}
  \label{eq:5}
  u_l=e^{- \kappa l}
\end{equation}
which turns out to be an eigenvector for $h>J$, with
\begin{equation}
  \label{eq:6}
  \kappa = \ln \frac h J
\end{equation}
and
\begin{equation}
  \label{eq:7}
  E= 2 J \cosh \kappa = h+\frac{J^2}h.
\end{equation}
We thus find a critical field value, above which one energy eigenstate
splits off from the quasi-continuum (\ref{eq:4}) and moves upward with
growing $h$. The localization length of that state is proportional to
$(\ln h/J)^{-1}$ and its amplitude at the boundary taking into account
normalization is given by
\begin{equation}
  \label{eq:8}
  (u_1)^2 = 1-\left(\frac J h \right)^{2}.
\end{equation}
The remaining eigenstates of the system are extended and similar to
(\ref{eq:3}); the appropriate Ansatz is
\begin{equation}
  \label{eq:9}
  {u_l= \sin (ql + \delta),}
\end{equation}
{where $q$ is a wave vector between 0 and $\pi$. For finite $N$
the value of $q$ is fixed by the boundary condition at the far end of
the system, leading to  $q=\nu \pi /(N+1)$ for $h=0$ as in
Eq. (\ref{eq:3}), while for $N \to \infty$ $q$ becomes continuous and can take any value between 0 and $\pi$. The
state (\ref{eq:9}) yields the eigenvalue}
\begin{equation}
  \label{eq:10}
  E= 2 J \cos q
\end{equation}
{and the phase shift $\delta$ obeys}
\begin{equation}
  \label{eq:11}
  \tan \delta = \frac {h \sin q}{J-h \cos q}.
\end{equation}
Obviously $\delta$ has a singularity for $h>J$ at some
$q$. However, that singularity is only a jump from $\pi/2$ to $-\pi/2$
(or vice versa) corresponding to an irrelevant sign change in the
eigenvector components.

If $N$ is finite (but large) the amplitudes of all extended
eigenstates scale as $N^{-1/2}$, while the amplitude (\ref{eq:8}) of
the localized state at the boundary site is constant and approaches
unity as $h$ grows. This is the key to the (quasi-) PST reported in
Ref.\ \onlinecite{CLMS09}.
As $N$ or $h$ grow the localized states at the boundaries will
  increasingly dominate the transfer of localized quantum information
  between the ends of the chain, and we expect  the speed of that
  transfer to be proportional to the overlap between the states
  localized at the left and right ends of the chain, respectively,
  which by (\ref{eq:5}) and (\ref{eq:6}) is proportional to
  $e^{-\kappa N} = \left(\frac Jh\right)^N$.

\section{The finite chain: approximate results and perfect state transfer}
\label{three}

We now consider the full model (\ref{eq:1}), with a strong boundary
magnetic field $h \gg J$ at both ends of the $N$-site chain. We then
have two extremely localized energy eigenstates with energies far
above all other states. The system can thus be approximately described
by two isolated spins in a strong field and an $(N-2)$-site chain in
between. The corresponding Hamiltonian matrix is obtained from
(\ref{eq:2}) by setting
\begin{equation}
  \label{eq:12}
  H_{12}=H_{21}=H_{N,N-1}=H_{N-1,N}=0 .
\end{equation}
Taking parity into account, the two localized states are given by the
amplitudes 
\begin{equation}
  \label{eq:13}
  u^{\pm}_1=\pm u^{\pm}_N = \frac 1{\sqrt 2} ; \;  u^{\pm}_l = 0
  \mbox{ otherwise}
\end{equation}
both states have energy $E_{\pm}=h$. The $N-2$ extended states  are given by
\begin{equation}
  \label{eq:13b}
  u_1^{\nu}=u_N^{\nu}=0; \; u_l^{\nu}= 
\left(
\frac 2{N-1}
\right)^{\frac 12}
\sin \frac {\nu \pi (l-1)}{N-1}
\end{equation}
$(\nu=1,...,N-2)$, with energy eigenvalues
\begin{equation}
  \label{eq:13c}
    E_{\nu} = 2 J \cos \frac {\nu \pi }{N-1}.
\end{equation}

Returning to the full problem defined by the Hamiltonian matrix
(\ref{eq:2}) we expect modified energy eigenvalues $E_{\pm}$ for the
formerly degenerate isolated boundary states $\ket{\pm}$. The initial and 
final states of the desired perfect state transfer can be written in terms
of the states (\ref{eq:13}):
\begin{equation}
  \label{eq:14}
  \ket{1}=\frac 1{\sqrt2} \left( \ket{+} + \ket{-} \right)  \quad ; \quad 
 \ket{N}=\frac 1{\sqrt2} \left( \ket{+} - \ket{-} \right)  
\end{equation}
and consequently the transfer time $\tau$ is equal to the time during
which a phase difference $\pi$ develops between the energy eigenstates
$\ket{+}$ and $\ket{-}$, that is
\begin{equation}
  \label{eq:15}
  \tau = \frac{\pi}{E_+ - E_-} .
\end{equation}
The energy difference ${E_+ - E_-}$ can be 
determined approximately by treating the matrix elements $J$ between
sites 1 and 2 and sites $N-1$ and $N$, respectively, as perturbations.
{The corresponding perturbation operator $V$ then is a $N \times
  N$ matrix with four nonzero entries $J$. The matrix elements of $V$
  between the unperturbed eigenstates (\ref{eq:13}) vanish and thus
  first-order degenerate perturbation theory does not lift the
  degeneracy. In second order, the corrections to the energy are given
  \cite{LL65} by the eigenvalues of the matrix
  \begin{equation}
    \label{eq:15a}
    \sum_{\nu \neq n,n^{\prime}}
    \frac{\langle n |V|\nu \rangle \langle \nu|V|n^{\prime}\rangle}{E_n^{(0)}-E_{\nu}^{(0)}},
  \end{equation}
where $|n\rangle$ and $|n^{\prime}\rangle$ are  the unperturbed degenerate
energy eigenstates with eigenvalue $E_n^{(0)}=h$, and $|\nu \rangle$
runs over all other unperturbed eigenstates (\ref{eq:13b}), with eigenvalues
$E_{\nu}^{(0)}$ (\ref{eq:13c}).
In our case the degenerate subspace is spanned by $|+\rangle$ and $|-\rangle$, or equivalently, by $|1\rangle$ and $|N\rangle$, compare (\ref{eq:14}). The matrix elements of the perturbation in the basis $\{ |1\rangle,|N\rangle\}$  are (compare  (\ref{eq:13b}))
\begin{equation}
  \label{eq:15b}
  \langle 1 |V|\nu \rangle = Ju_2^{\nu} = J \left(\frac 2{N-1} \right)^{\frac 1 2} \sin \frac{\nu \pi}{N-1}
\end{equation}
and
\begin{equation}
  \label{eq:15c}
    \langle N |V|\nu \rangle = Ju_{N-1}^{\nu} = (-1)^{\nu+1}   \langle 1 |V|\nu \rangle .
\end{equation}
The diagonal elements $(n=n^{\prime})$ of the 2 $\times$ 2 matrix (\ref{eq:15a}) thus turn out to be equal, leading to a mere energy shift. The non-diagonal elements are both  equal to
\begin{equation}
  \label{eq:15d}
  \sum_{\nu=1}^{N-2} 
\frac{ \langle 1 |V|\nu \rangle  \langle \nu |V|N \rangle}{h-2J \cos \frac{\nu \pi}{N-1}}
= \frac{2J^2}{N-1}  \sum_{\nu=1}^{N-2}
\frac{ (-1)^{\nu+1} \sin^2 \frac{\nu \pi}{N-1}}{h-2J \cos \frac{\nu \pi}{N-1}},
\end{equation}
leading to an energy splitting equal to twice the non-diagonal element (\ref{eq:15d}):
}
\begin{equation}
  \label{eq:16}
  E_+ - E_- = \frac{4J}{N-1} \sum_{\nu=1}^{N-2} \frac{(-1)^{\nu+1} \sin^2 
\frac{\pi \nu}{N-1}}{\frac h J - 2 \cos \frac{\pi \nu}{N-1}}.
\end{equation}
\begin{widetext}
An asymptotic expansion of (\ref{eq:16}) in the small parameter $J/h$ reads
\begin{equation}
  \label{eq:17}
   E_+ - E_- = 2J \left( \frac Jh \right)^{N-2} \left[ 1+ (N-3) \left(
       \frac Jh \right)^2 + \frac{(N^2-3N-2)}2 \left( \frac Jh
     \right)^4 + {\cal O} \left(\left( \frac Jh \right)^6\right)\right].
\end{equation}
\end{widetext}
This confirms the behavior which was to be expected based on the
results of Sec. \ref{two} and it explains
the exponential growth of the transfer time (\ref{eq:15})
with both $N$ and $h/J$. 

The exact expression of the energy difference $E_+-E_-$
can be found and approximated by
the following simple consideration. The boundary-localized symmetric
and antisymmetric eigenstates are determined using an ansatz of the
form 
\begin{equation}
  \label{eq:18}
  u^{\pm}_l = e^{-\kappa l} \pm e^{-\kappa(N+1-l)}.
\end{equation}
The corresponding energy eigenvalues are again given by 
\begin{equation}
  \label{eq:19}
  E_{\pm}=2J \cosh \kappa_{\pm},
\end{equation}
(compare Eq.(\ref{eq:7})) but now the values $\kappa_{\pm}$ fulfill
the transcendental equation 
\begin{equation}
e^{\kappa_\pm}
= \frac{h}{J}\pm e^{-N\kappa_\pm}
\left(\frac{h}{J}e^{\kappa_\pm}-1\right)
\end{equation}
which results from the condition for
the eigenvalue at sites 1 or $N$.
This equation can be solved
iteratively, starting from $\kappa_0=\ln \frac hJ$ (compare
Eq.(\ref{eq:6})). To lowest order in  $(J/h)^{N-2}$ this yields
\begin{equation}
  \label{eq:20}
  E_+-E_- =  2J \Big( \frac Jh \Big)^{N-2}
\Big(1-\Big( \frac Jh \Big)^2\Big)+ 
{\cal O}\Big( \Big( \frac Jh \Big)^{2N-4}\Big).
\end{equation}

For comparison to the numerical results of Ref.\
\onlinecite{CLMS09} we identify $h$ with $\Delta_E$ from that reference and
set $J=2$.

%%%%%%%%%%%%%%%%%%%%%%%%%%%%%%%%%%%%%%%%%%%%%%%%%%%%%%%%%%%%%%%%%%%%%%%%%
\begin{table}[ht] 
  \begin{tabular}{c|c|c|c|c}
 
 $h$ & Ref.\ \onlinecite{CLMS09} & exact diag.\ & perturbative & order 
$(J/h)^{N-2}$ \\
\hline
10 & 99 & 107 & 90  &107\\
20 &8010&801& 770 & 801\\
30 & 2665 & 2674& 2627 & 2674\\
40&6260&6315&6252 & 6315\\
50&12294&12311&12233 & 12311\\

  \end{tabular}

\caption{Transfer times for a chain of length $N=5$ with different values 
of the boundary field $h$ as obtained by various methods. See text for details.}
\label{un} 
\end{table}
%%%%%%%%%%%%%%%%%%%%%%%%%%%%%%%%%%%%%%%%%%%%%%%%%%%%%%%%%%%%%%%%%%%%%%%%%

In Table \ref{un} we compare the transfer time for $N=5$ as obtained
by different methods and for different values of $h$. The column
labelled ``Ref. \onlinecite{CLMS09}'' contains the numbers reported in
that reference (with an obvious typing error for $h=20$). The next
column shows the transfer time (\ref{eq:15}) from the difference of the 
two dominant eigenvalues $E_+$ and $E_-$ as obtained from an exact numerical
diagonalization of the Hamiltonian (\ref{eq:2}), whereas the fourth
column is obtained from the perturbation calculation described
above. 

The asymptotic expansion (\ref{eq:17}) agrees to the expression (\ref{eq:16}) 
to a much higher precision than that of the numbers in the table. 
However, it can be seen that the perturbative result approaches the 
exact one only for fairly large values of $h$. On the other hand,
the lowest order in $\left( \frac Jh \right)^{N-2}$, given in Eq.\ (\ref{eq:20}),
 agrees with the exact values to all digits given (compare the last column in Table \ref{un}).

We have performed similar comparisons for chain lengths
$N=7$ and $N=12$, obtaining similar results, that is, for sufficiently
strong boundary fields the perturbation calculation yields the
transfer time to a precision of one percent or better while the
lowest order fits the exact result within $10^{-7}$ for $N=7$
and within $10^{-14}$ for $N=12$, as was to be expected from Eq.\ (\ref{eq:20}).
However, at
$N=12$ the difference between the two leading energy eigenvalues
approaches the precision limits of standard numerical procedures and
the corresponding transfer times become prohibitively long by any
standards.

{Finally, we briefly discuss the degree of fidelity which can be
  reached by the scheme discussed here. A reasonable measure of the
  transfer fidelity between the states $|1\rangle$ and $|N\rangle$ is 
  \begin{equation}
    \label{eq:25}
    f_{1N}(t) = |\langle 1 | e^{-iHt}| N \rangle|^2.
  \end{equation}
Employing the energy eigenstates $|\nu \rangle$ and the eigenvalues
$E_{\nu}$ we can write 
\begin{eqnarray}
  \label{eq:26}
  \langle 1 | e^{-iHt}| N \rangle &=& 
\sum_{\nu} \langle 1|\nu\rangle   \langle \nu | N \rangle e^{-iE_{\nu}t}\\  \nonumber
&=&  \langle 1|+\rangle   \langle + | N \rangle e^{-iE_{+}t}\\
\nonumber
& & +
 \langle 1|-\rangle   \langle - | N \rangle e^{-iE_{-}t} \\ \nonumber
& & + \sum_{\mu} \langle 1|\mu\rangle   \langle \mu | N \rangle e^{-iE_{\mu}t},
\end{eqnarray}
where we have separated the dominant (localized) eigenstates
$|\pm\rangle$ from the remainig (extended) eigenstates $|\mu\rangle$,
each of which yields a contribution ${\cal O}(1/N)$, with a
quasi-random phase. The transfer time $\tau$ is the instant when 
the two dominant terms in (\ref{eq:26}) interfere constructively,
maximizing $f_{1N}$ and thus
leading to (compare (\ref{eq:8}))
\begin{equation}
  \label{eq:27}
   |\langle 1 | e^{-iH\tau}| N \rangle| \approx 1-\left( \frac{J}{h}  \right)^2
\end{equation}
or
\begin{equation}
  \label{eq:28}
  f_{1N}(\tau) = 1 -2\left( \frac{J}{h}  \right)^2.
\end{equation}
The dominant time
dependence of $ f_{1N}(t)$ is a slow oscillation between zero and the
maximum value (\ref{eq:28}) with period $2\tau$, with superimposed
rapid small-amplitude oscillations from the subdominant terms in
(\ref{eq:26}). This is precisely what numerical calculations for small
$N$ show.
}

{The value  $f_{1N}(\tau)$ does not depend on the chain length $N$ and it grows
(increasingly slowly) as $h$ grows. Note, however, that the transfer
time $\tau$ grows exponentially as $h$ grows. For a given $N$ the
choice of $h$ thus is the result of a trade-off between transfer time (\ref{eq:15})
and fidelity (\ref{eq:28}), depending on the relative importance assigned to each of
these two figures of merit. Hence, our analytical results for transfer time and fidelity provide the basis for finding an optimum trade-off in practice.}

\begin{acknowledgments}
JS wishes to thank Analia Zwick for helpful discussions. GSU is
supported by the DFG through grant UH 90/5-1.
\end{acknowledgments}

\bibliographystyle{apsrev}
\bibliography{/home/stolze/BIBLIO/jsbas_def,/home/stolze/BIBLIO/general,/home/stolze/BIBLIO/mypapers,/home/stolze/BIBLIO/qcomp}

\newcommand{\noopsort}[1]{}
\begin{thebibliography}{23}
\expandafter\ifx\csname natexlab\endcsname\relax\def\natexlab#1{#1}\fi
\expandafter\ifx\csname bibnamefont\endcsname\relax
  \def\bibnamefont#1{#1}\fi
\expandafter\ifx\csname bibfnamefont\endcsname\relax
  \def\bibfnamefont#1{#1}\fi
\expandafter\ifx\csname citenamefont\endcsname\relax
  \def\citenamefont#1{#1}\fi
\expandafter\ifx\csname url\endcsname\relax
  \def\url#1{\texttt{#1}}\fi
\expandafter\ifx\csname urlprefix\endcsname\relax\def\urlprefix{URL }\fi
\providecommand{\bibinfo}[2]{#2}
\providecommand{\eprint}[2][]{\url{#2}}

\bibitem[{\citenamefont{Bose}(2007)}]{Bos07}
\bibinfo{author}{\bibfnamefont{S.}~\bibnamefont{Bose}},
  \bibinfo{journal}{Contemp. Phys.} \textbf{\bibinfo{volume}{48}},
  \bibinfo{pages}{13} (\bibinfo{year}{2007}).

\bibitem[{\citenamefont{Kay}(2010)}]{Kay10}
\bibinfo{author}{\bibfnamefont{A.}~\bibnamefont{Kay}}, \bibinfo{journal}{Int.
  J. Quant. Inf} \textbf{\bibinfo{volume}{8}}, \bibinfo{pages}{641}
  (\bibinfo{year}{2010}).

\bibitem[{\citenamefont{Osborne and Linden}(2004)}]{OL04}
\bibinfo{author}{\bibfnamefont{T.~J.} \bibnamefont{Osborne}} \bibnamefont{and}
  \bibinfo{author}{\bibfnamefont{N.}~\bibnamefont{Linden}},
  \bibinfo{journal}{Phys. Rev. A} \textbf{\bibinfo{volume}{69}},
  \bibinfo{pages}{052315} (\bibinfo{year}{2004}).

\bibitem[{\citenamefont{Christandl et~al.}(2004)\citenamefont{Christandl,
  Datta, Ekert, and Landahl}}]{CDEL04}
\bibinfo{author}{\bibfnamefont{M.}~\bibnamefont{Christandl}},
  \bibinfo{author}{\bibfnamefont{N.}~\bibnamefont{Datta}},
  \bibinfo{author}{\bibfnamefont{A.}~\bibnamefont{Ekert}}, \bibnamefont{and}
  \bibinfo{author}{\bibfnamefont{A.~J.} \bibnamefont{Landahl}},
  \bibinfo{journal}{Phys. Rev. Lett.} \textbf{\bibinfo{volume}{92}},
  \bibinfo{pages}{187902} (\bibinfo{year}{2004}).

\bibitem[{\citenamefont{Albanese et~al.}(2004)\citenamefont{Albanese,
  Christandl, Datta, and Ekert}}]{ACDE04}
\bibinfo{author}{\bibfnamefont{C.}~\bibnamefont{Albanese}},
  \bibinfo{author}{\bibfnamefont{M.}~\bibnamefont{Christandl}},
  \bibinfo{author}{\bibfnamefont{N.}~\bibnamefont{Datta}}, \bibnamefont{and}
  \bibinfo{author}{\bibfnamefont{A.}~\bibnamefont{Ekert}},
  \bibinfo{journal}{Phys. Rev. Lett.} \textbf{\bibinfo{volume}{93}},
  \bibinfo{pages}{230502} (\bibinfo{year}{2004}).

\bibitem[{\citenamefont{Yung and Bose}(2005)}]{YB04}
\bibinfo{author}{\bibfnamefont{M.-H.} \bibnamefont{Yung}} \bibnamefont{and}
  \bibinfo{author}{\bibfnamefont{S.}~\bibnamefont{Bose}},
  \bibinfo{journal}{Phys. Rev. A} \textbf{\bibinfo{volume}{71}},
  \bibinfo{pages}{032310} (\bibinfo{year}{2005}).

\bibitem[{\citenamefont{Shi et~al.}(2005)\citenamefont{Shi, Li, Song, and
  Sun}}]{SLSS04}
\bibinfo{author}{\bibfnamefont{T.}~\bibnamefont{Shi}},
  \bibinfo{author}{\bibfnamefont{Y.}~\bibnamefont{Li}},
  \bibinfo{author}{\bibfnamefont{Z.}~\bibnamefont{Song}}, \bibnamefont{and}
  \bibinfo{author}{\bibfnamefont{C.-P.} \bibnamefont{Sun}},
  \bibinfo{journal}{Phys. Rev. A} \textbf{\bibinfo{volume}{71}},
  \bibinfo{pages}{032309} (\bibinfo{year}{2005}).

\bibitem[{\citenamefont{Karbach and Stolze}(2005)}]{my44}
\bibinfo{author}{\bibfnamefont{P.}~\bibnamefont{Karbach}} \bibnamefont{and}
  \bibinfo{author}{\bibfnamefont{J.}~\bibnamefont{Stolze}},
  \bibinfo{journal}{Phys. Rev. A} \textbf{\bibinfo{volume}{72}},
  \bibinfo{pages}{030301(R)} (\bibinfo{year}{2005}).

\bibitem[{\citenamefont{Wang et~al.}(2011)\citenamefont{Wang, Shuang, and
  Rabitz}}]{WSR11}
\bibinfo{author}{\bibfnamefont{Y.}~\bibnamefont{Wang}},
  \bibinfo{author}{\bibfnamefont{F.}~\bibnamefont{Shuang}}, \bibnamefont{and}
  \bibinfo{author}{\bibfnamefont{H.}~\bibnamefont{Rabitz}},
  \bibinfo{journal}{Phys.\ Rev.\ A} \textbf{\bibinfo{volume}{84}},
  \bibinfo{pages}{012307} (\bibinfo{year}{2011}).

\bibitem[{\citenamefont{W\'{o}jcik et~al.}(2005)\citenamefont{W\'{o}jcik,
  {\L}uczak, Kurzy\'{n}ski, Grudka, Gdala, and Bednarska}}]{WLKx05}
\bibinfo{author}{\bibfnamefont{A.}~\bibnamefont{W\'{o}jcik}},
  \bibinfo{author}{\bibfnamefont{T.}~\bibnamefont{{\L}uczak}},
  \bibinfo{author}{\bibfnamefont{P.}~\bibnamefont{Kurzy\'{n}ski}},
  \bibinfo{author}{\bibfnamefont{A.}~\bibnamefont{Grudka}},
  \bibinfo{author}{\bibfnamefont{T.}~\bibnamefont{Gdala}}, \bibnamefont{and}
  \bibinfo{author}{\bibfnamefont{M.}~\bibnamefont{Bednarska}},
  \bibinfo{journal}{Phys.\ Rev.\ A} \textbf{\bibinfo{volume}{72}},
  \bibinfo{pages}{034303} (\bibinfo{year}{2005}).

\bibitem[{\citenamefont{Banchi et~al.}(2010)\citenamefont{Banchi, Apollaro,
  Cuccoli, Vaia, and Verrucchi}}]{BACx10}
\bibinfo{author}{\bibfnamefont{L.}~\bibnamefont{Banchi}},
  \bibinfo{author}{\bibfnamefont{T.~J.~G.} \bibnamefont{Apollaro}},
  \bibinfo{author}{\bibfnamefont{A.}~\bibnamefont{Cuccoli}},
  \bibinfo{author}{\bibfnamefont{R.}~\bibnamefont{Vaia}}, \bibnamefont{and}
  \bibinfo{author}{\bibfnamefont{P.}~\bibnamefont{Verrucchi}},
  \bibinfo{journal}{Phys.\ Rev.\ A} \textbf{\bibinfo{volume}{82}},
  \bibinfo{pages}{052321} (\bibinfo{year}{2010}).

\bibitem[{\citenamefont{Banchi et~al.}(2011)\citenamefont{Banchi, Apollaro,
  Cuccoli, Vaia, and Verrucchi}}]{BACx11}
\bibinfo{author}{\bibfnamefont{L.}~\bibnamefont{Banchi}},
  \bibinfo{author}{\bibfnamefont{T.~J.~G.} \bibnamefont{Apollaro}},
  \bibinfo{author}{\bibfnamefont{A.}~\bibnamefont{Cuccoli}},
  \bibinfo{author}{\bibfnamefont{R.}~\bibnamefont{Vaia}}, \bibnamefont{and}
  \bibinfo{author}{\bibfnamefont{P.}~\bibnamefont{Verrucchi}},
  \bibinfo{journal}{New J. Phys.} \textbf{\bibinfo{volume}{13}},
  \bibinfo{pages}{123006} (\bibinfo{year}{2011}).

\bibitem[{\citenamefont{Zwick and Osenda}(2011)}]{ZO11}
\bibinfo{author}{\bibfnamefont{A.}~\bibnamefont{Zwick}} \bibnamefont{and}
  \bibinfo{author}{\bibfnamefont{O.}~\bibnamefont{Osenda}},
  \bibinfo{journal}{J.\ Phys.\ A: Math.\ Gen.} \textbf{\bibinfo{volume}{44}},
  \bibinfo{pages}{105302} (\bibinfo{year}{2011}).

\bibitem[{\citenamefont{Oh et~al.}(2011)\citenamefont{Oh, Wu, Shim, Fei,
  Friesen, and Hu}}]{OWSx11}
\bibinfo{author}{\bibfnamefont{S.}~\bibnamefont{Oh}},
  \bibinfo{author}{\bibfnamefont{L.-A.} \bibnamefont{Wu}},
  \bibinfo{author}{\bibfnamefont{Y.-P.} \bibnamefont{Shim}},
  \bibinfo{author}{\bibfnamefont{J.}~\bibnamefont{Fei}},
  \bibinfo{author}{\bibfnamefont{M.}~\bibnamefont{Friesen}}, \bibnamefont{and}
  \bibinfo{author}{\bibfnamefont{X.}~\bibnamefont{Hu}},
  \bibinfo{journal}{Phys.\ Rev.\ A} \textbf{\bibinfo{volume}{84}},
  \bibinfo{pages}{022330} (\bibinfo{year}{2011}).

\bibitem[{\citenamefont{Yao et~al.}(2011)\citenamefont{Yao, Jiang, Gorshkov,
  Gong, Zhai, Duan, and Lukin}}]{YJGx11}
\bibinfo{author}{\bibfnamefont{N.}~\bibnamefont{Yao}},
  \bibinfo{author}{\bibfnamefont{L.}~\bibnamefont{Jiang}},
  \bibinfo{author}{\bibfnamefont{A.~V.} \bibnamefont{Gorshkov}},
  \bibinfo{author}{\bibfnamefont{Z.-X.} \bibnamefont{Gong}},
  \bibinfo{author}{\bibfnamefont{A.}~\bibnamefont{Zhai}},
  \bibinfo{author}{\bibfnamefont{L.-M.} \bibnamefont{Duan}}, \bibnamefont{and}
  \bibinfo{author}{\bibfnamefont{M.~D.} \bibnamefont{Lukin}},
  \bibinfo{journal}{Phys.\ Rev.\ Lett.} \textbf{\bibinfo{volume}{106}},
  \bibinfo{pages}{040505} (\bibinfo{year}{2011}).

\bibitem[{\citenamefont{Casaccino et~al.}(2009)\citenamefont{Casaccino, Lloyd,
  Mancini, and Severini}}]{CLMS09}
\bibinfo{author}{\bibfnamefont{A.}~\bibnamefont{Casaccino}},
  \bibinfo{author}{\bibfnamefont{S.}~\bibnamefont{Lloyd}},
  \bibinfo{author}{\bibfnamefont{S.}~\bibnamefont{Mancini}}, \bibnamefont{and}
  \bibinfo{author}{\bibfnamefont{S.}~\bibnamefont{Severini}},
  \bibinfo{journal}{Int. J. Quant. Inf.} \textbf{\bibinfo{volume}{7}},
  \bibinfo{pages}{1417} (\bibinfo{year}{2009}).

\bibitem[{\citenamefont{Zwick et~al.}(2011)\citenamefont{Zwick, \'Alvarez,
  Stolze, and Osenda}}]{my60}
\bibinfo{author}{\bibfnamefont{A.}~\bibnamefont{Zwick}},
  \bibinfo{author}{\bibfnamefont{G.~A.} \bibnamefont{\'Alvarez}},
  \bibinfo{author}{\bibfnamefont{J.}~\bibnamefont{Stolze}}, \bibnamefont{and}
  \bibinfo{author}{\bibfnamefont{O.}~\bibnamefont{Osenda}},
  \bibinfo{journal}{Phys. Rev. A} \textbf{\bibinfo{volume}{84}},
  \bibinfo{pages}{022311} (\bibinfo{year}{2011}).

\bibitem[{\citenamefont{Zwick et~al.}(2012)\citenamefont{Zwick, \'Alvarez,
  Stolze, and Osenda}}]{my62}
\bibinfo{author}{\bibfnamefont{A.}~\bibnamefont{Zwick}},
  \bibinfo{author}{\bibfnamefont{G.~A.} \bibnamefont{\'Alvarez}},
  \bibinfo{author}{\bibfnamefont{J.}~\bibnamefont{Stolze}}, \bibnamefont{and}
  \bibinfo{author}{\bibfnamefont{O.}~\bibnamefont{Osenda}},
  \bibinfo{journal}{Phys. Rev. A} \textbf{\bibinfo{volume}{85}},
  \bibinfo{pages}{012318} (\bibinfo{year}{2012}).

\bibitem[{\citenamefont{G\"uttinger}(1932)}]{Gue32}
\bibinfo{author}{\bibfnamefont{P.}~\bibnamefont{G\"uttinger}},
  \bibinfo{journal}{Z. Physik} \textbf{\bibinfo{volume}{73}},
  \bibinfo{pages}{169} (\bibinfo{year}{1932}).

\bibitem[{\citenamefont{Pauli}(1933)}]{Pau33}
\bibinfo{author}{\bibfnamefont{W.}~\bibnamefont{Pauli}},
  \emph{\bibinfo{title}{Handbuch der Physik}} (\bibinfo{publisher}{Springer,
  Berlin}, \bibinfo{year}{1933}), \bibinfo{note}{p. 162}.

\bibitem[{\citenamefont{Hellmann}(1937)}]{Hel37}
\bibinfo{author}{\bibfnamefont{H.}~\bibnamefont{Hellmann}},
  \emph{\bibinfo{title}{Einf\"uhrung in die Quantenchemie}}
  (\bibinfo{publisher}{Franz Deuticke, Leipzig}, \bibinfo{year}{1937}),
  \bibinfo{note}{p. 285}.

\bibitem[{\citenamefont{Feynman}(1939)}]{Fey39}
\bibinfo{author}{\bibfnamefont{R.~P.} \bibnamefont{Feynman}},
  \bibinfo{journal}{Phys. Rev.} \textbf{\bibinfo{volume}{56}},
  \bibinfo{pages}{340} (\bibinfo{year}{1939}).

\bibitem[{\citenamefont{Landau and Lifshitz}(1965)}]{LL65}
\bibinfo{author}{\bibfnamefont{L.}~\bibnamefont{Landau}} \bibnamefont{and}
  \bibinfo{author}{\bibfnamefont{E.}~\bibnamefont{Lifshitz}},
  \emph{\bibinfo{title}{Quantum Mechanics}} (\bibinfo{publisher}{Pergamon
  Press, Oxford}, \bibinfo{year}{1965}), \bibinfo{edition}{2nd} ed.

\end{thebibliography}

\end{document}